\documentclass[aps,prl,twocolumn,showpacs,groupedaddress]{revtex4}  
\usepackage{graphicx}  
\usepackage{dcolumn}   
\usepackage{bm}        
\usepackage{amssymb}   

\newcommand{\ttb}{$t\bar{t}$}
\newcommand{\bbb}{$b\bar{b}$}

\newcommand{\ppb}{$p\bar{p}$}

\newcommand{\invpb}{pb$^{-1}$}
\newcommand{\invfb}{fb$^{-1}$}

\newcommand{\met}{\mbox{\ensuremath{\,\slash\kern-.7emE_{T}}}}

\newcommand{\mht}{\mbox{\ensuremath{\,\slash\kern-.7emH_{T}}}}

\def\lsim{\mathrel{\rlap{\lower4pt\hbox{$\sim$}}
    \raise1pt\hbox{$<$}}}                

\newcommand{\HT}{H_T}

\newcommand{\etadet}{\vert\eta_{\mathrm{det}}\vert}

\newcommand{\W}{W}
\newcommand{\Z}{Z}

\newcommand{\sq}{\tilde{q}}
\newcommand{\sg}{\tilde{g}}

\newcommand{\xo}{\tilde{{\chi}}^0_1}

\begin{document}

\hspace{5.2in} \mbox{Fermilab-Pub-07/668-E}

\title{Search for squarks and gluinos in events with jets and missing transverse energy\\
using 2.1\,fb$^{\bm{-1}}$ of $\bm{p\bar{p}}$ collision data at $\bm{\sqrt{s}=}$1.96\,TeV
}
%
\author{V.M.~Abazov$^{36}$}
\author{B.~Abbott$^{76}$}
\author{M.~Abolins$^{66}$}
\author{B.S.~Acharya$^{29}$}
\author{M.~Adams$^{52}$}
\author{T.~Adams$^{50}$}
\author{E.~Aguilo$^{6}$}
\author{S.H.~Ahn$^{31}$}
\author{M.~Ahsan$^{60}$}
\author{G.D.~Alexeev$^{36}$}
% ...

\affiliation{\vspace{0.1 in}(The D\O\ Collaboration)\vspace{0.1 in}}
% ...
\affiliation{$^{6}$University of Alberta, Edmonton, Alberta, Canada,
                Simon Fraser University, Burnaby, British Columbia, Canada,
                York University, Toronto, Ontario, Canada, and
                McGill University, Montreal, Quebec, Canada}
% ...
\affiliation{$^{29}$Tata Institute of Fundamental Research, Mumbai, India}
% ...
\affiliation{$^{31}$Korea Detector Laboratory, Korea University, Seoul, Korea}
% ...
\affiliation{$^{36}$Joint Institute for Nuclear Research, Dubna, Russia}
% ...
\affiliation{$^{50}$Florida State University, Tallahassee, Florida 32306, USA}
% ...
\affiliation{$^{52}$University of Illinois at Chicago,
                Chicago, Illinois 60607, USA}
% ...
\affiliation{$^{60}$Kansas State University, Manhattan, Kansas 66506, USA}
% ...
\affiliation{$^{66}$Michigan State University,
                East Lansing, Michigan 48824, USA}
% ...
\affiliation{$^{76}$University of Oklahoma, Norman, Oklahoma 73019, USA}
% ...
\date{January 24, 2008}

\begin{abstract}
A data sample corresponding to an integrated luminosity of 2.1\,\invfb\
collected by the D0 detector at the Fermilab Tevatron Collider was
analyzed to search for squarks and gluinos produced in \ppb\ collisions
at a center-of-mass energy of 1.96~TeV. No evidence for the production
of such particles was observed in topologies involving jets and missing
transverse energy, and 95\% C.L. lower limits of 379~GeV and 308~GeV
were set on the squark and gluino masses, respectively, within the 
framework of minimal supergravity with $\tan\beta = 3$, $A_0 = 0$, and $\mu < 0$.
The corresponding previous limits are improved by 54~GeV and 67~GeV.
\end{abstract}

\maketitle 

In supersymmetric models~\cite{susy}, each of the standard model (SM) particles has a partner
differing by a half-unit of spin. If $R$-parity~\cite{rparity} is conserved, supersymmetric 
particles are produced in pairs, and their decay leads to SM particles and to the lightest 
supersymmetric particle (LSP), which is stable. Cosmological arguments suggest that
the LSP should be neutral and colorless~\cite{cosmo}. The lightest neutralino $\xo$, which is a 
mixture of the superpartners of the neutral gauge and Higgs bosons, fulfills these conditions.
In the following, it is assumed that $R$-parity is conserved and that $\xo$ is the LSP. Since
this particle is weakly interacting, it escapes detection and provides the classic missing 
transverse energy ($\met$) signature at colliders.
In $p\bar{p}$ collisions, squarks ($\sq$) and gluinos ($\sg$), the superpartners of 
quarks and gluons, would be abundantly produced, if sufficiently light, by the strong interaction,
leading to final states containing jets and $\met$. 
The most constraining direct limits on squark and gluino masses were obtained by the D0
collaboration~\cite{Abazov:2006bj}, based on an analysis of 310\,\invpb\ of data from 
$p\bar{p}$ collisions at a center-of-mass energy of 1.96\,TeV, collected during Run~II of 
the Fermilab Tevatron Collider. 
In the model of minimal supergravity (mSUGRA)~\cite{msugra}, the limits obtained were 
$m_{\sq}>325$~GeV and $m_{\sg}>241$~GeV at the 95\% C.L.,
for the set of model parameters detailed below.
In this Letter, an update of the D0 search for squarks and gluinos in topologies with 
jets and large $\met$ is reported, using a seven times larger data set of 2.1\,\invfb.

A detailed description of the D0 detector can be found in Ref.~\cite{Abazov:2005pn}.
The central tracking system consists of a silicon microstrip tracker and a central fiber tracker,
both located within a 2~T superconducting solenoidal magnet. A liquid-argon 
and uranium calorimeter covers pseudorapidities up to $|\eta|$ $\approx 4.2$, where 
$\eta=-\ln \left[ \tan \left( \theta/2 \right) \right]$, and $\theta$ is the polar angle 
with respect to the proton beam direction. The calorimeter consists of three sections, 
housed in separate cryostats: the central one covers $|\eta|$ $\lsim 1.1$, and the two end 
sections extend the coverage to larger $\vert\eta\vert$. The calorimeter is segmented in depth, 
with four electromagnetic layers followed by up to five hadronic layers. It is also segmented 
in semi-projective towers of size $0.1\times 0.1$ in the $(\eta,\phi)$ plane, 
where $\phi$ is the azimuthal 
angle in radians. Calorimeter cells are defined by the intersections of towers and layers. Additional 
sampling is provided by scintillating tiles
between cryostats.
An outer muon system, covering $|\eta|<2$, consists of a layer of tracking detectors and scintillation 
trigger counters in front of 1.8~T iron toroids, followed by two similar layers
after the toroids. 
Jets were reconstructed with the iterative midpoint cone algorithm~\cite{jetalgo} with cone radius
$\mathcal{R}= \sqrt{(\Delta\phi)^2+(\Delta y)^2}=0.5$ in azimuthal angle $\phi$ and rapidity
$y= \frac{1}{2} \ln ((E+p_{z})/(E-p_{z}))$.
The jet energy scale (JES) was derived from the transverse momentum balance in photon-plus-jet 
events. The $\met$ was calculated from all calorimeter cells, and corrected for the jet energy 
scale and for the transverse momenta of reconstructed muons.

The D0 trigger system consists of three levels, L1, L2 and L3. The 1.2\,\invfb\ of data recorded
after 2006 (during the so-called Run~IIb) were collected using a significantly upgraded system, in particular
a new L1 calorimeter trigger \cite{Abolins:2007yz} involving a sliding-window algorithm which 
improved the jet triggering efficiency. In addition, $\met$ was used at L1 to select events,
which was not done during the previous data taking period (Run~IIa).
Both improvements helped to keep a high trigger efficiency despite the 
increased instantaneous luminosity in Run~IIb (up to $2.8\times 10^{32}\rm{cm}^{-2}\rm{s}^{-1}$, to 
be compared to $1.6\times 10^{32}\rm{cm}^{-2}\rm{s}^{-1}$ in Run~IIa).
The events used in this analysis were recorded using two categories of triggers~\cite{thomas}.
The dijet triggers selected events with two jets and $\met$, while the multijet triggers
were optimized for events with at least three jets and $\met$. 

The SM processes leading to events with jets and real $\met$ in the final state (``SM 
backgrounds'') are
the production of $W$ or $Z$ bosons in association with jets ($W/Z+$jets), of pairs of vector 
bosons ($WW$, $WZ$, $ZZ$) or top quarks ($t\bar{t}$), and of single top quarks.
The neutrinos from
the decays $Z\to\nu\bar{\nu}$ and $W\to l\nu$, with the $W$ boson produced directly or coming from 
a top quark decay, generate the $\met$ signature.
In this analysis, most of the $W$ boson leptonic decays leading to an electron or a muon were
identified, and the corresponding events rejected. However, a charged lepton from $W$ boson decay
can be a tau decaying hadronically. 
It can also be an electron or a muon that escapes detection or fails the identification 
criteria. Such $W+$jets events therefore exhibit the jets plus $\met$ signature.
Finally, multijet production also leads to a final state 
with jets and $\met$ when one or more jets are mismeasured (``QCD background'').

Events from SM and supersymmetric processes were simulated using Monte Carlo (MC) generators and passed
through a full {\sc geant3}-based~\cite{geant} simulation of the detector geometry and response. They were
subsequently processed with the same reconstruction chain as the data. The parton density functions (PDFs)
used in the MC generators are the {\sc CTEQ6L1}~\cite{cteq6} PDFs. A data event from a randomly selected
beam crossing was overlaid on each event to simulate the additional minimum bias interactions. 
The QCD background was not simulated, but estimated directly from data. 
The {\sc alpgen} generator~\cite{Mangano:2002ea} was used to simulate $W/Z+$\,jets and $t\bar{t}$
production.
It was interfaced with {\sc pythia}~\cite{Sjostrand:2006za} for the simulation of initial
and final state radiation (ISR/FSR) and of jet hadronization. Pairs of vector bosons and electroweak top
quark production were simulated with {\sc pythia} and {\sc comphep}~\cite{Boos:2004kh}, respectively.
The next-to-leading order (NLO) cross sections were computed with {\sc mcfm\,5.1}~\cite{Campbell:2001ik}.

Squark and gluino production and decay were simulated with {\sc pythia}. The masses of the supersymmetric
particles were calculated with {\sc suspect\,2.3}~\cite{Djouadi:2002ze} from the set of five
mSUGRA parameters: $m_0$, the universal scalar mass at the grand unification scale $\Lambda_{\rm{GUT}}$;
$m_{1/2}$ the universal gaugino mass at $\Lambda_{\rm{GUT}}$; $A_0$, the universal trilinear coupling at
$\Lambda_{\rm{GUT}}$; $\tan\beta$, the ratio of the vacuum expectation values of the two Higgs fields; and 
$\mu$ the sign of the Higgs-mixing mass parameter. The decay widths and branching ratios
of all supersymmetric particles were then
calculated with {\sc sdecay\,1.1a}~\cite{Muhlleitner:2003vg}.
In order to allow for an easier comparison with previous results, the
following parameters were fixed to the same values as in Ref.~\cite{Abazov:2006bj}:
$A_0 = 0$, $\tan\beta = 3$, and $\mu < 0$. The production of scalar top quarks (stops) was ignored, 
and in the following, the ``squark mass'' is the average mass of all squarks other than stops.
All squark and gluino decay modes were taken into account, including
cascade decays through charginos or neutralinos which could additionally produce electrons,
muons, or taus. The NLO cross sections of the squark and gluino pair production
were calculated with {\sc prospino\,2}~\cite{Beenakker:1996ch}.

The analysis strategy is the same as in Ref.~\cite{Abazov:2006bj} with three analyses optimized 
for three benchmark regions of the mSUGRA parameter space. A ``dijet'' analysis was optimized at low
$m_0$ for events containing a pair of acoplanar jets, as expected from
$p\bar p\to\sq\overline{\sq} \to q\xo\bar q\xo$ and 
$p\bar p\to\sq\sq \to q\xo q\xo$.
A ``gluino'' analysis was optimized at high $m_0$ for events with at least four jets, as expected from
$p\bar p\to \sg\sg \to q\bar q\xo q\bar q\xo$. Finally, a ``3-jets''
analysis was optimized for events with at least three jets, as expected from
$p\bar p\to\sq\sg\to q\xo q\bar q\xo$. The benchmark for this
analysis is the case where $m_{\sq}=m_{\sg}$.

Each analysis required at least two jets and substantial $\met$ ($\geq$ 40 GeV).
The acoplanarity, i.e. the azimuthal angle between the two leading jets, 
was required to be smaller than $165^\circ$, where
the two leading jets are those with the largest transverse momenta.  
These leading jets were required: to be in the central region of the
calorimeter, $\etadet < 0.8$, where $\eta_{\mathrm{det}}$ is the jet
pseudorapidity calculated under the assumption that the jet originates
from the detector center; to have fractions of energy in the
electromagnetic layers of the calorimeter less than 0.95; and to have
transverse momenta greater than 35\,GeV.
The best primary vertex (PV0) was selected among all reconstructed primary vertices
(PV) as the one with the smallest probability to be due to a minimum bias interaction~\cite{vtxreco}.
Its longitudinal position with respect to the detector center was restricted to
$\vert z\vert < 60$\,cm to ensure efficient reconstruction. 
For each jet, CPF0 is defined as the fraction of track $p_T$ sum associated with the jet which comes
from PV0,
CPF0$= \sum p^{{\rm track}}_T ({\rm PV0}) / \sum p^{{\rm track}}_T ({\rm any\ PV})$.
The two leading jets were required to have CPF0 larger than 0.75. 
Since jet transverse momenta and $\met$ were calculated with respect to PV0, this criterion reduced the
background with large $\met$ due to an incorrect PV0 selection.

\begin{table}
\caption{\label{cutflow1}
Selection criteria for the three analyses (all energies and momenta in GeV); see the text 
for further details.
}
\begin{ruledtabular}
\begin{tabular}{cccc}
Preselection Cut & \multicolumn{3}{c}{All Analyses} \\
\hline
$\met$				        & \multicolumn{3}{c}{$\geq 40$}			\\
$|\mathrm{Vertex}\ z\ {\rm pos.}|$	& \multicolumn{3}{c}{$<60$ cm}			\\
Acoplanarity			        & \multicolumn{3}{c}{$< 165^\circ$}		\\
\hline
Selection Cut			& ``dijet''	& ``3-jets''	& ``gluino'' 		\\
\hline
Trigger				& dijet		& multijet	& multijet		\\
${\rm jet}_1$ $p_T$\footnotemark[1]     & $\geq 35$ 	& $\geq 35$ 	& $\geq 35$		\\
${\rm jet}_2$ $p_T$\footnotemark[1]	& $\geq 35$ 	& $\geq 35$ 	& $\geq 35$		\\
${\rm jet}_3$ $p_T$\footnotemark[2]	& $-$		& $\geq 35$ 	& $\geq 35$		\\
${\rm jet}_4$ $p_T$\footnotemark[2]	& $-$		& $-$		& $\geq 20$		\\
\hline
Electron veto			& yes		& yes		& yes			\\
Muon veto			& yes		& yes		& yes			\\
\hline
$\Delta\phi (\met,\mathrm{jet_1})$ 		& $\geq 90^\circ$ & $\geq 90^\circ$ & $\geq 90^\circ$ 	\\
$\Delta\phi (\met,\mathrm{jet_2})$ 		& $\geq 50^\circ$ & $\geq 50^\circ$ & $\geq 50^\circ$ 	\\
$\Delta\phi_{\rm{min}}(\met,\mathrm{any\,jet})$ & $\geq 40^\circ$	& $-$ & $-$				\\
\hline
$\HT$				& $\geq$ 325  & $\geq$ 375  & $\geq$ 400 		\\
$\met$				& $\geq$ 225  & $\geq$ 175  & $\geq$ 100 		\\
\end{tabular}
\end{ruledtabular}
\footnotetext[1]{First and second jets are also required to be central \mbox{($\etadet<0.8$)}, 
with an electromagnetic fraction below 0.95, and to have \mbox{$\rm{CPF0} \geq 0.75$}.}
\footnotetext[2]{Third and fourth jets are required to have \mbox{$\etadet<2.5$}, 
with an electromagnetic fraction below 0.95.}
\end{table}

Different selection criteria were applied in the three analyses, as summarized in
Table~\ref{cutflow1}. Events passing a trigger for acoplanar dijet events were used in 
the ``dijet'' analysis, while a multijet and $\met$ trigger was required in the ``3-jets''
and ``gluino'' analyses. In the ``3-jets'' and ``gluino'' analyses, a third and fourth 
jet were required, respectively. In comparison with the previously published
analysis~\cite{Abazov:2006bj}, the $\etadet$ upper limit 
for the third and fourth jet was increased from 0.8 to 2.5, which considerably improved
the signal efficiency at high $m_0$. These two jets also had to have their
fraction of energy in the electromagnetic layers of the calorimeter less than 0.95, but no
requirement was made on their CPF0. The third and fourth jet transverse momenta were required to 
exceed 35\,GeV and 20\,GeV, respectively.
In all three analyses, a veto on isolated electrons or muons 
with $p_T>10$\,GeV was applied to reject background events containing a leptonic $W$ boson decay.
The azimuthal angles between the $\met$ and the first jet, 
$\Delta\phi(\met,\mathrm{jet_1})$, and the second jet, 
$\Delta\phi(\met,\mathrm{jet_2})$, were used to remove events where the energy of 
one jet was mismeasured, generating $\met$ aligned with that jet.
In the ``dijet'' analysis, the minimum azimuthal angle $\Delta\phi_{\rm{min}}(\met,\mathrm{any\,jet})$
between the $\met$ and any jet with $p_T>15$\,GeV was required to be greater than $40^{\circ}$ to further
suppress QCD background. This criterion was not used in the ``3-jets'' and ``gluino'' analyses because of the
higher jet multiplicity in the signal events.

\begin{figure*}
\begin{tabular}{ccc}
\includegraphics[width=5.5cm]{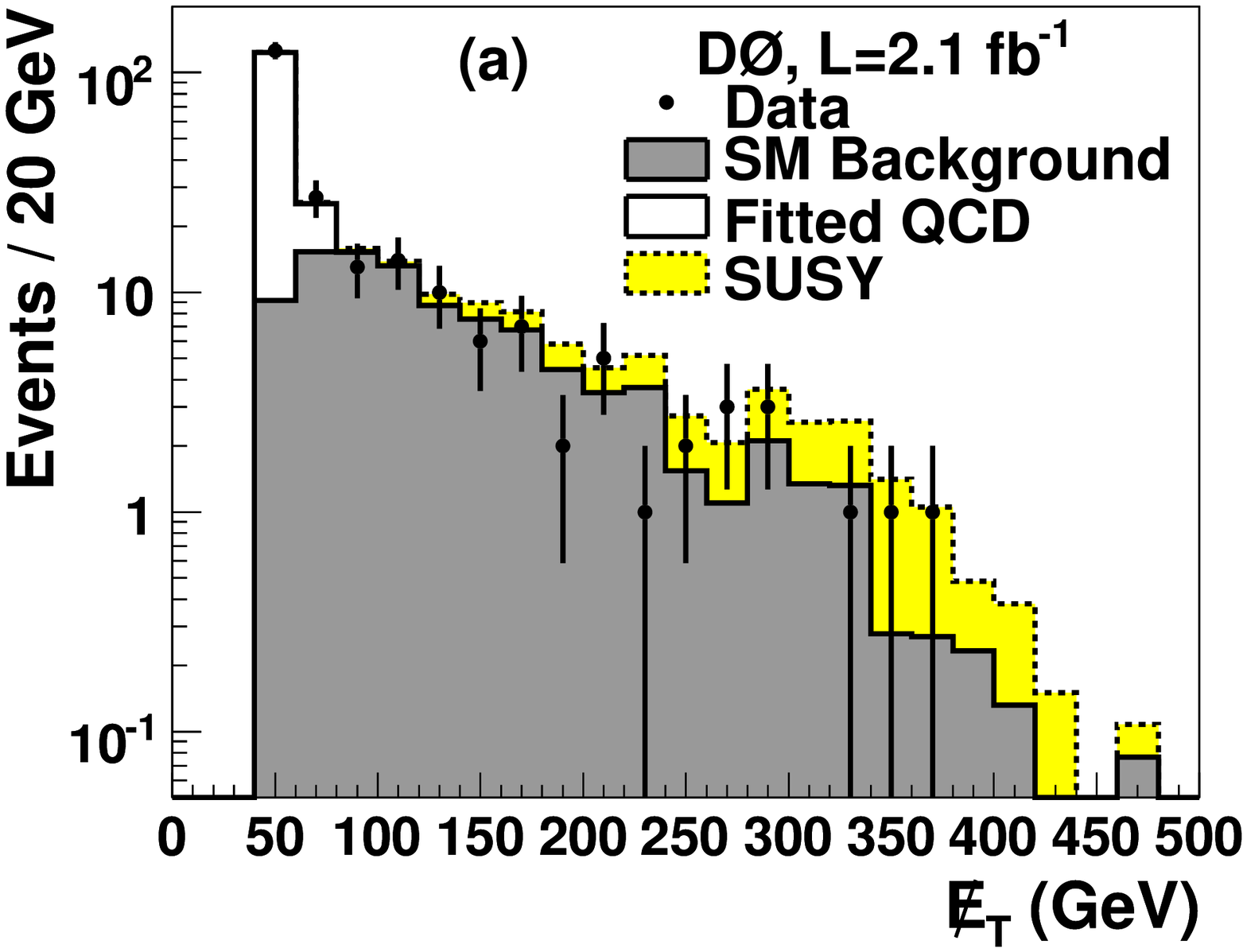} &
\includegraphics[width=5.5cm]{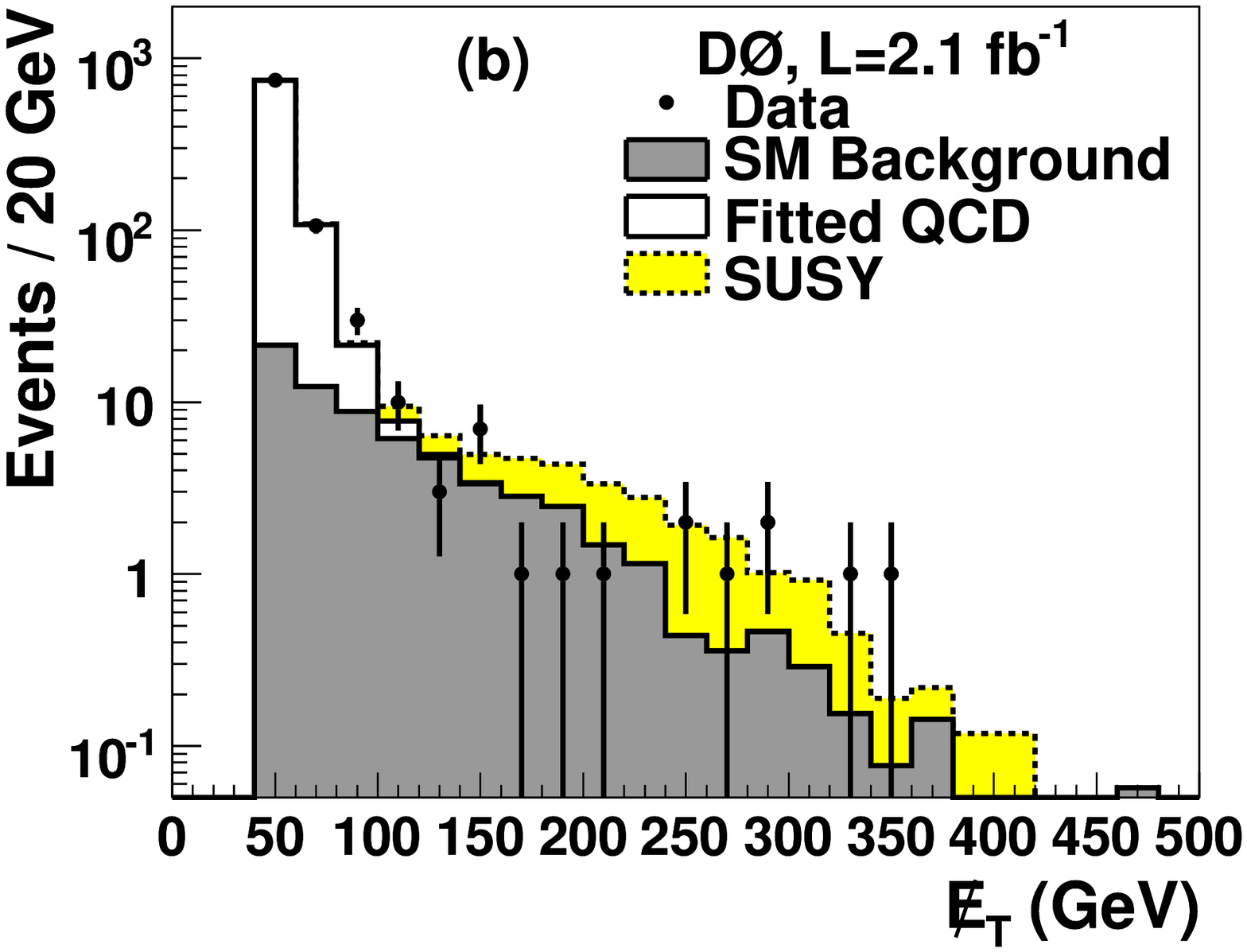} &
\includegraphics[width=5.5cm]{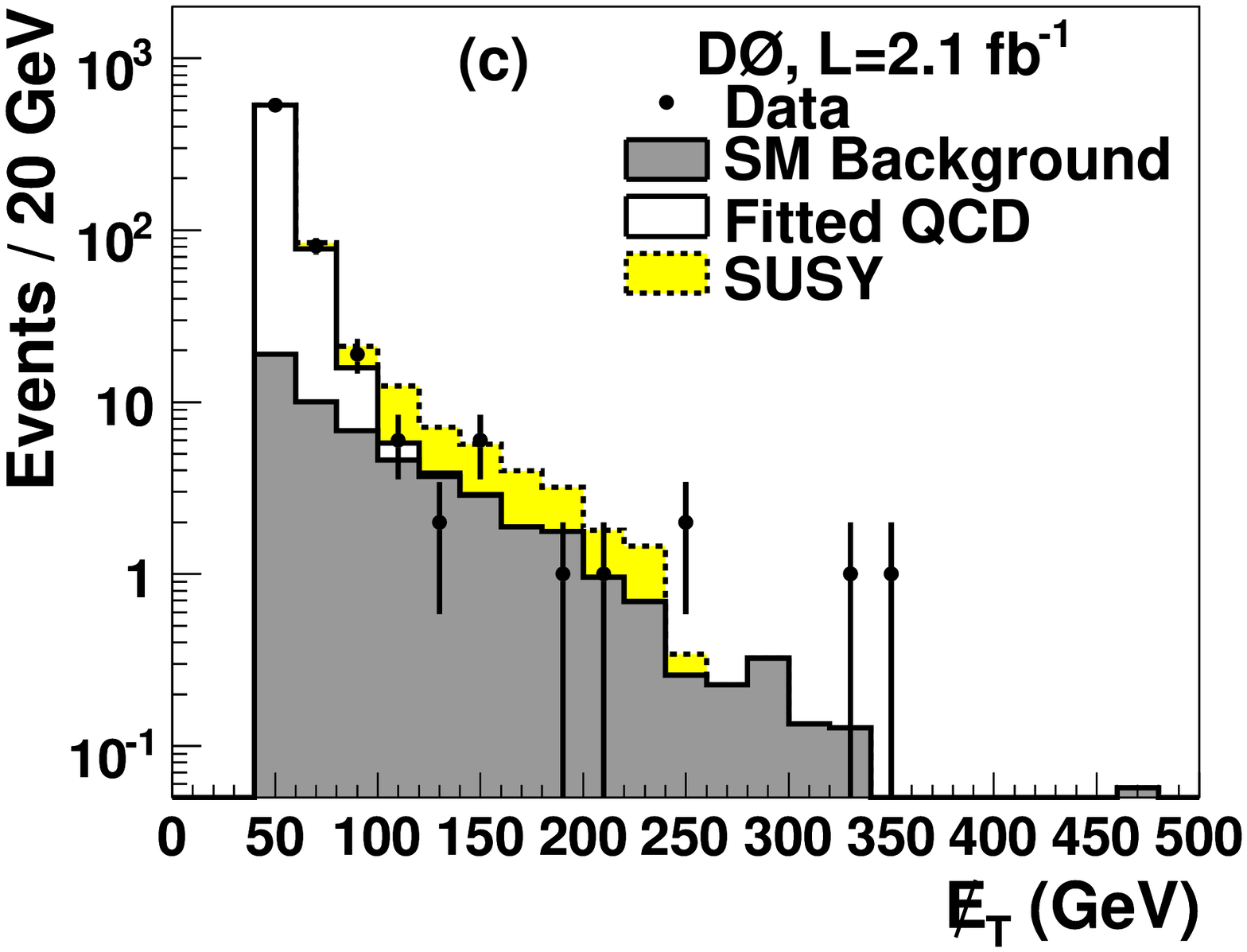} \\
\end{tabular}
\caption{\label{plots}
Distributions of $\met$ after applying all analysis criteria except the one on $\met$ for the 
``dijet'' (a), ``3-jets'' (b), and ``gluino'' (c) analyses;
for data (points with error bars), for SM background (full histogram), 
and for signal MC (dotted histogram on top of SM). 
The signal drawn corresponds to $(m_0,m_{1/2})=(25,175)$\,GeV, $m_{\sq}=m_{\sg}=410$\,GeV,
and $(m_0,m_{1/2})=(500,110)$\,GeV from left to right.
The fitted QCD background contribution is also shown.
} 
\end{figure*}

The two final cuts on $\met$ and on $\HT = \sum_{\mathrm{jets}} p_T$, where the sum is over all jets
with $p_T>15$\,GeV and $\etadet<2.5$, were optimized by minimizing 
the expected upper limit on the cross section in the absence of signal. Here and in the calculation of
the final limits, the modified frequentist {\sl CL}$_s$ method~\cite{CLS} was used. The QCD background
contribution was estimated by fitting the $\met$ distribution below 60\,GeV with an exponential function,
after subtraction of the SM background processes, and subsequently extrapolating this function 
above the chosen $\met$ cut value. The optimal cuts are given in Table~\ref{cutflow1}. 
The $\met$ distributions after applying all analysis criteria except the one on
$\met$ are shown in Fig.~\ref{plots} for the three analyses.

\begin{table*}
\renewcommand{\arraystretch}{1.2}
\begin{center}
\caption{\label{summary1}
For each analysis, information on the signal for which it was optimized 
($m_0$, $m_{1/2}$, $m_{\sg}$, $m_{\sq}$, and nominal NLO cross section),
signal efficiency, the number of events observed, the number of events 
expected from SM backgrounds, the number of events expected from signal,
and the 95\% C.L. signal cross section upper limit.
The first uncertainty is statistical and the second is systematic.
}
\begin{ruledtabular}
\begin{tabular}{lcccccccc}
Analysis   & $(m_0,m_{1/2})$ & $(m_{\sg},m_{\sq})$ & $\sigma_{\mathrm{nom}}$ & $\epsilon_{\mathrm{sig.}}$     & $N_{\mathrm{obs.}}$ & $N_{\mathrm{backgrd.}}$ & $N_{\mathrm{sig.}}$ & $\sigma_{95}$\\
           & (GeV)           & (GeV)               & (pb)                   & (\%)                             &            	    &                         &		& (pb) \\
\hline
``dijet''  &  (25,175)       & (439,396)           & 0.072  	             & $6.8 \pm 0.4 ^{+1.2}_{-1.2}$ & 11             & $11.1 \pm 1.2 ^{+2.9}_{-2.3}$ & $ 10.4 \pm 0.6 ^{+1.8}_{-1.8}$ & 0.075 \\
``3-jets'' & (197,154)       & (400,400)           & 0.083  	             & $6.8 \pm 0.4 ^{+1.4}_{-1.3}$ &  9             & $10.7 \pm 0.9 ^{+3.1}_{-2.1}$ & $ 12.0 \pm 0.7 ^{+2.5}_{-2.3}$ & 0.065 \\
``gluino'' & (500,110)       & (320,551)           & 0.195  	             & $4.1 \pm 0.3 ^{+0.8}_{-0.7}$ & 20             & $17.7 \pm 1.1 ^{+5.5}_{-3.3}$ & $ 17.0 \pm 1.2 ^{+3.3}_{-2.9}$ & 0.165 \\
\end{tabular}
\end{ruledtabular}
\end{center}
\end{table*}

Table~\ref{summary1} reports the number of data events and the expected signal and background. 
The main background contributions are from
$(\Z\to\nu\bar{\nu})+\mathrm{jets}$, $(\W\to l\nu)+\mathrm{jets}$, and \ttb\,$\to$\,\bbb\,$q\bar{q}'l\nu$.
The QCD background was evaluated from a fit to the $\met$ distribution as described above.
The largest QCD contribution of $1.4\pm0.8$ event was estimated in the ``gluino'' analysis, but was conservatively
ignored when setting the limits. It was found to be negligible in the ``dijet'' and ``3-jets'' analyses.
The signal efficiencies are given in Table~\ref{summary1} for the three benchmark scenarios, with the
corresponding values of $m_0$, $m_{1/2}$, the squark and gluino masses, and the NLO cross section. 

\begin{table*}
\renewcommand{\arraystretch}{1.2}
\begin{center}
\caption{\label{rescombi}
Definition of the analysis combinations, and number of events observed in the data and expected 
from the SM backgrounds. 
}
\begin{ruledtabular}
\begin{tabular}{lcccrr}
Selection 		& ``dijet'' 	& ``3-jets'' 	& ``gluino'' 	& $N_{\rm{obs.}}$	& \multicolumn{1}{c}{$N_{\rm{backgrd.}}$} \\
\hline
Combination 1	& yes		& no		& no		&  8	& $ 9.4 \pm 1.2$ (stat.) $^{+2.3}_{-1.8}$ (syst.)\\
Combination 2	& no		& yes		& no		&  2	& $ 4.5 \pm 0.6$ (stat.) $^{+0.7}_{-0.5}$ (syst.)\\
Combination 3	& no		& no		& yes		& 14	& $12.5 \pm 0.9$ (stat.) $^{+3.6}_{-1.9}$ (syst.)\\
Combination 4	& yes		& yes		& no		&  1	& $ 1.1 \pm 0.3$ (stat.) $^{+0.5}_{-0.3}$ (syst.)\\
Combination 5	& yes		& no		& yes		&  \multicolumn{2}{r}{kinematically not allowed}    \\
Combination 6	& no		& yes		& yes		&  4	& $ 4.5 \pm 0.6$ (stat.) $^{+1.8}_{-1.3}$ (syst.)\\
Combination 7	& yes		& yes		& yes		&  2	& $ 0.6 \pm 0.2$ (stat.) $^{+0.1}_{-0.2}$ (syst.)\\
\hline
At least one selection	& \multicolumn{3}{c}{\ }			& 31	& $32.6 \pm 1.7$ (stat.) $^{+9.0}_{-5.8}$ (syst.)\\
\end{tabular}
\end{ruledtabular}
\end{center}
\end{table*}

The uncertainty coming from the JES corrections is typically (10--15)\% for the
SM backgrounds and (6--11)\% for the signal efficiencies.
The uncertainties due to the jet energy resolution, to the jet track confirmation, and to jet 
reconstruction and identification efficiencies range between 2\% and 4\%.
All these uncertainties on jet properties account for differences between
data and MC simulation, both for signal efficiencies and background contributions. 
The trigger was found
to be fully efficient for the event samples surviving all analysis cuts with an uncertainty of 2\%.
The uncertainty on the luminosity measurement is 6.1\%~\cite{d0lumi}.
All of these uncertainties are fully correlated between signal and SM backgrounds. 
A 15\% systematic uncertainty was set on the $W/Z$+jets and \ttb\ NLO cross sections.
The uncertainty on the signal acceptance due to the PDF choice was 
determined to be 6\%, using the forty-eigenvector basis of the {\sc CTEQ6.1M} PDF set~\cite{cteq6}.
Finally, the effects of ISR/FSR on the signal efficiencies were studied by varying
the {\sc pythia} parameters controlling the QCD scales and the maximal allowed
virtualities used in the simulation of the space-like and time-like
parton showers. The uncertainty on the signal efficiencies was determined to be 6\%.

\begin{figure*}
\begin{tabular}{ccc}
\includegraphics[width=5.5cm]{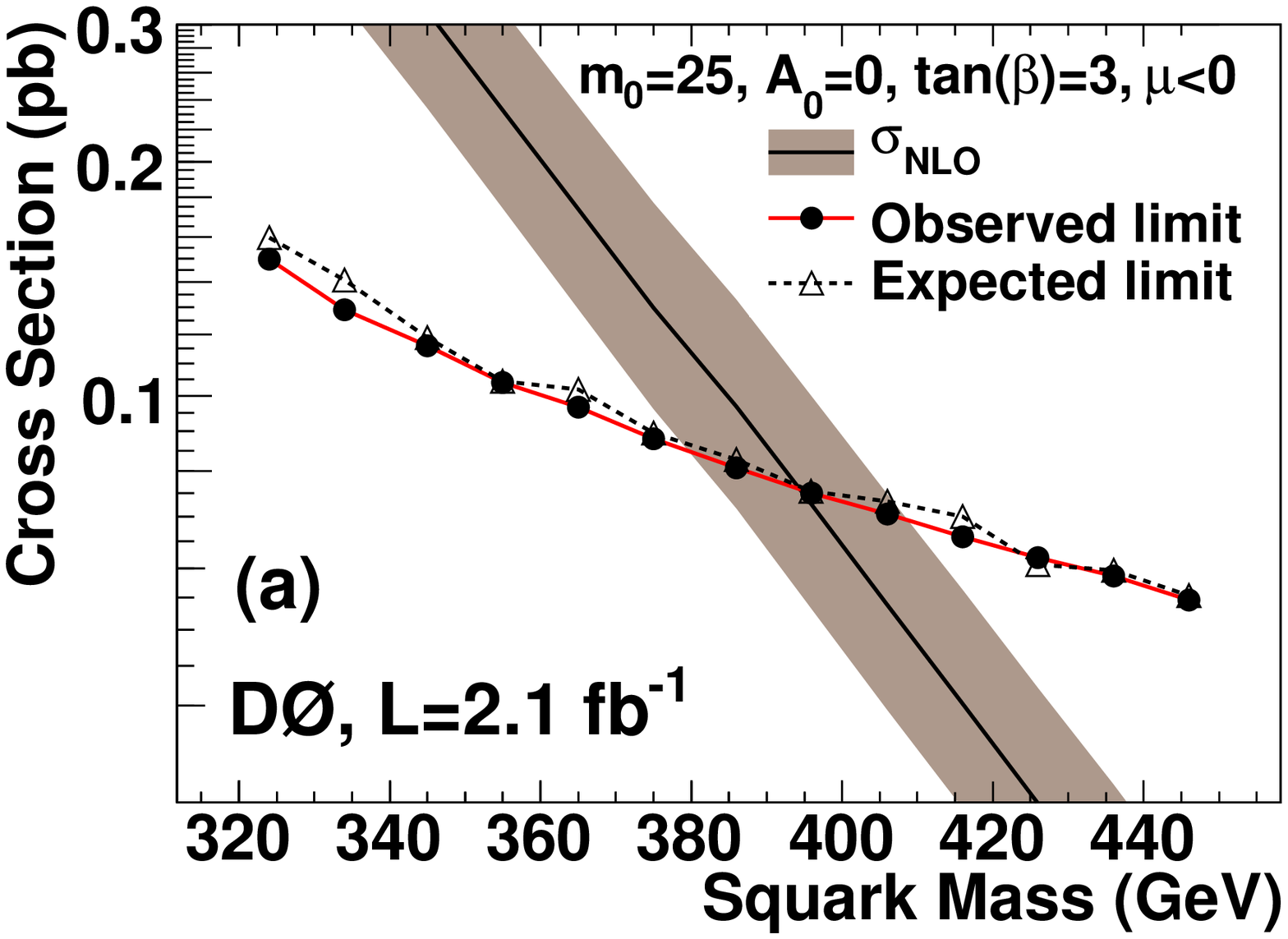} &
\includegraphics[width=5.5cm]{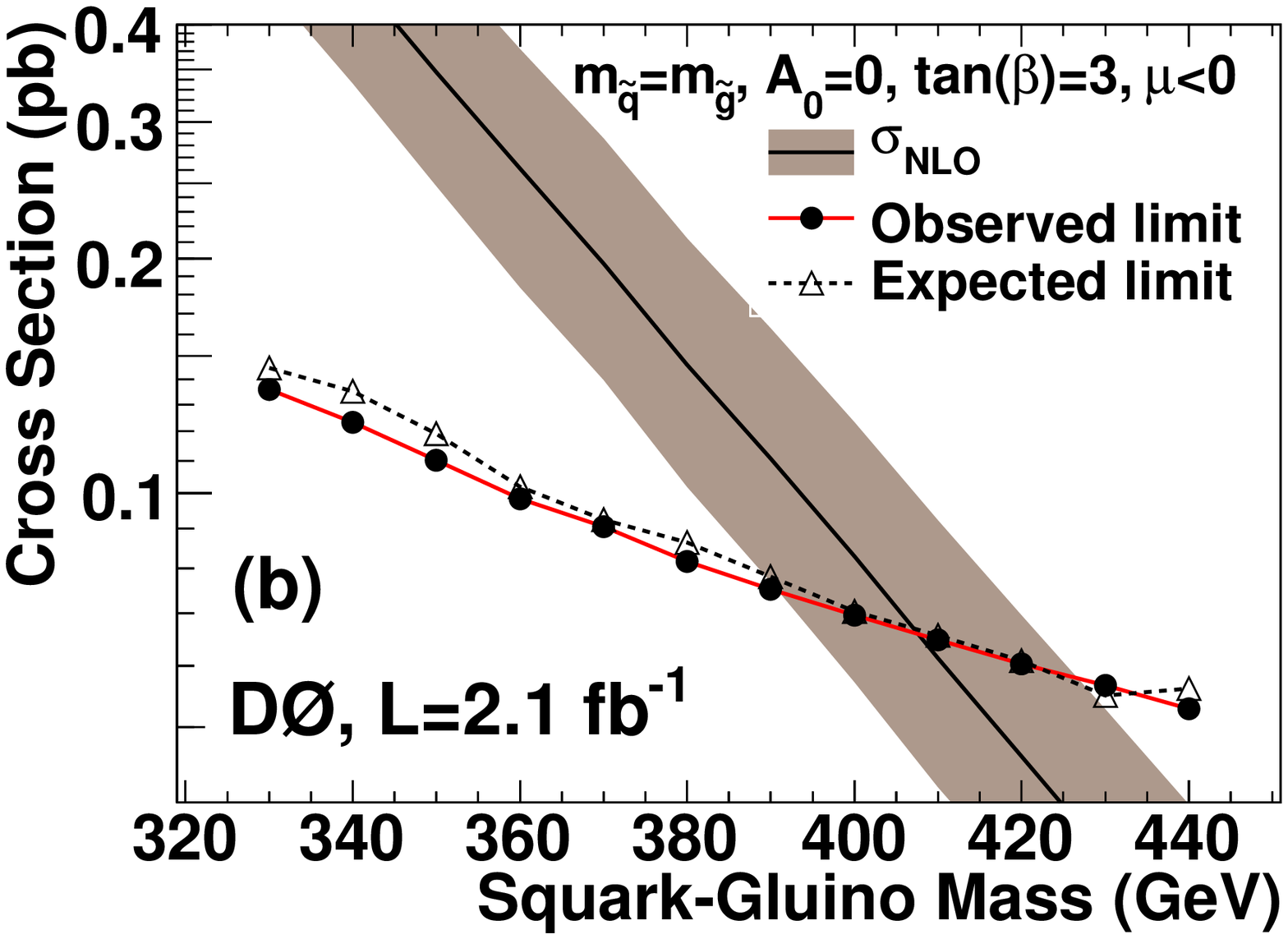} &
\includegraphics[width=5.5cm]{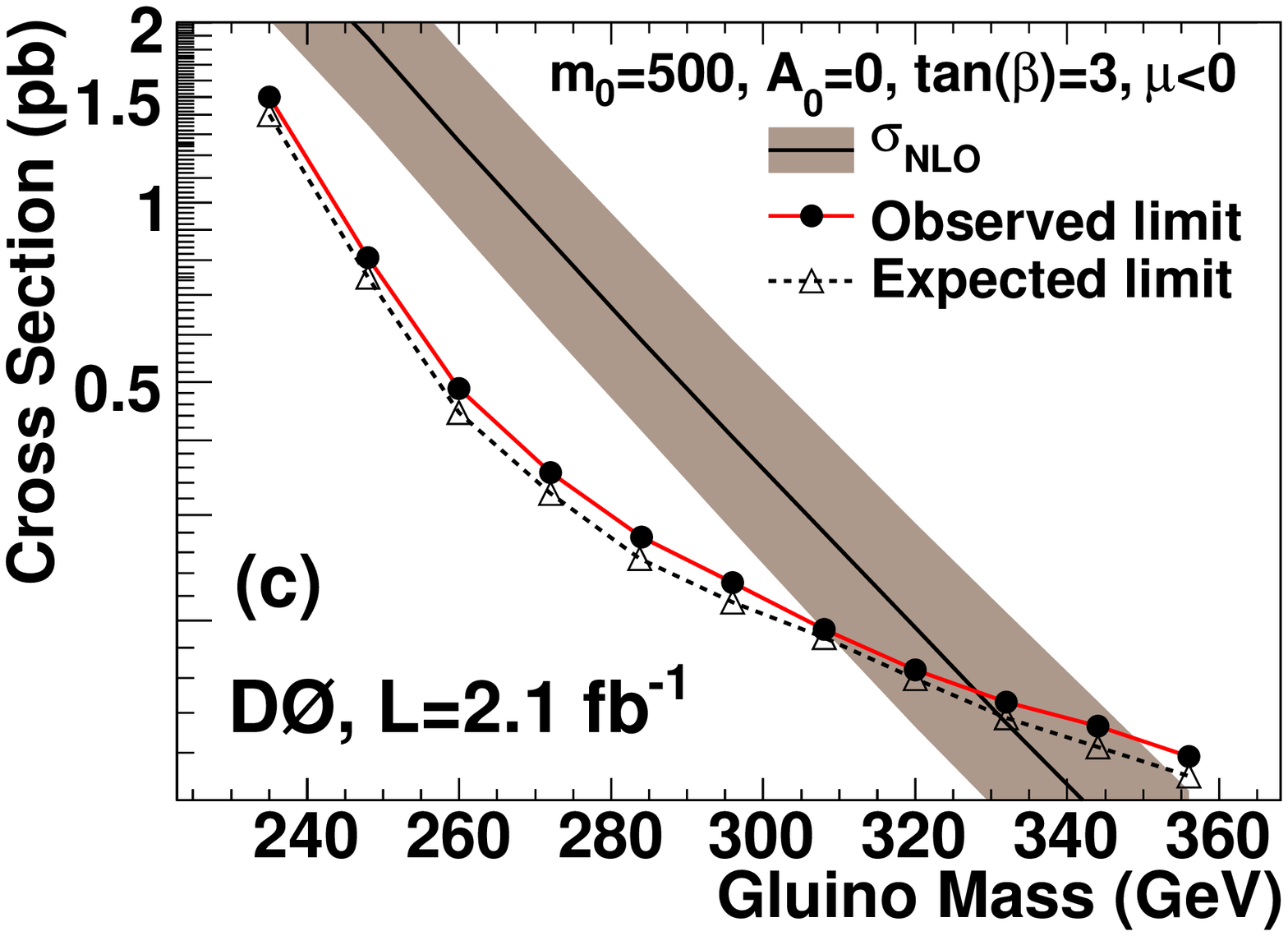} \\
\end{tabular}
\caption{\label{xseclim}
For $\tan\beta = 3$, $A_0 = 0$, $\mu < 0$, observed (closed circles) and expected (opened triangles) 
95\% C.L. upper limits on squark-gluino production cross sections combining the analyses for 
$m_0=25$\,GeV (a), $m_{\sq}=m_{\sg}$ (b), and $m_0=500$\,GeV (c). The nominal production 
cross sections are also shown, with shaded bands corresponding to the PDF and 
renormalization-and-factorization scale uncertainties.
} 
\end{figure*}

The nominal NLO signal cross sections, $\sigma_{\mathrm{nom}}$, were computed with the 
{\sc CTEQ6.1M} PDF and for the renormalization and factorization scale $\mu_{\mathrm{r,f}}=Q$, where $Q$ was taken to be equal 
to $m_{\sg}$ for $\sg\sg$ production, $m_{\sq}$ for $\sq\sq$ and $\sq\overline{\sq}$ production, 
and $(m_{\sq}+m_{\sg})/2$ for $\sq\sg$ production. 
The uncertainty due to the choice of PDF was determined using the full set
of {\sc CTEQ6.1M} eigenvectors, with the individual uncertainties added in
quadrature. The effect on the nominal signal cross sections, which varies between 15\% and 60\%, is
dominated by the large uncertainty on the gluon distribution at \mbox{high $x$}. 
The effect of the renormalization and factorization scale was studied by calculating the signal cross sections
for $\mu_{\mathrm{r,f}}=Q$, $\mu_{\mathrm{r,f}}=Q/2$ and $\mu_{\mathrm{r,f}}=2 \times Q$.
The factor of two on this scale reduces or increases the nominal signal cross sections by \mbox{(15--20)\%}.
The PDF and $\mu_{\mathrm{r,f}}$ effects were added in quadrature to compute minimum, 
$\sigma_{\mathrm{min}}$, and maximum, $\sigma_{\mathrm{max}}$, signal cross sections. 

\begin{figure}
\includegraphics[width=8.5cm]{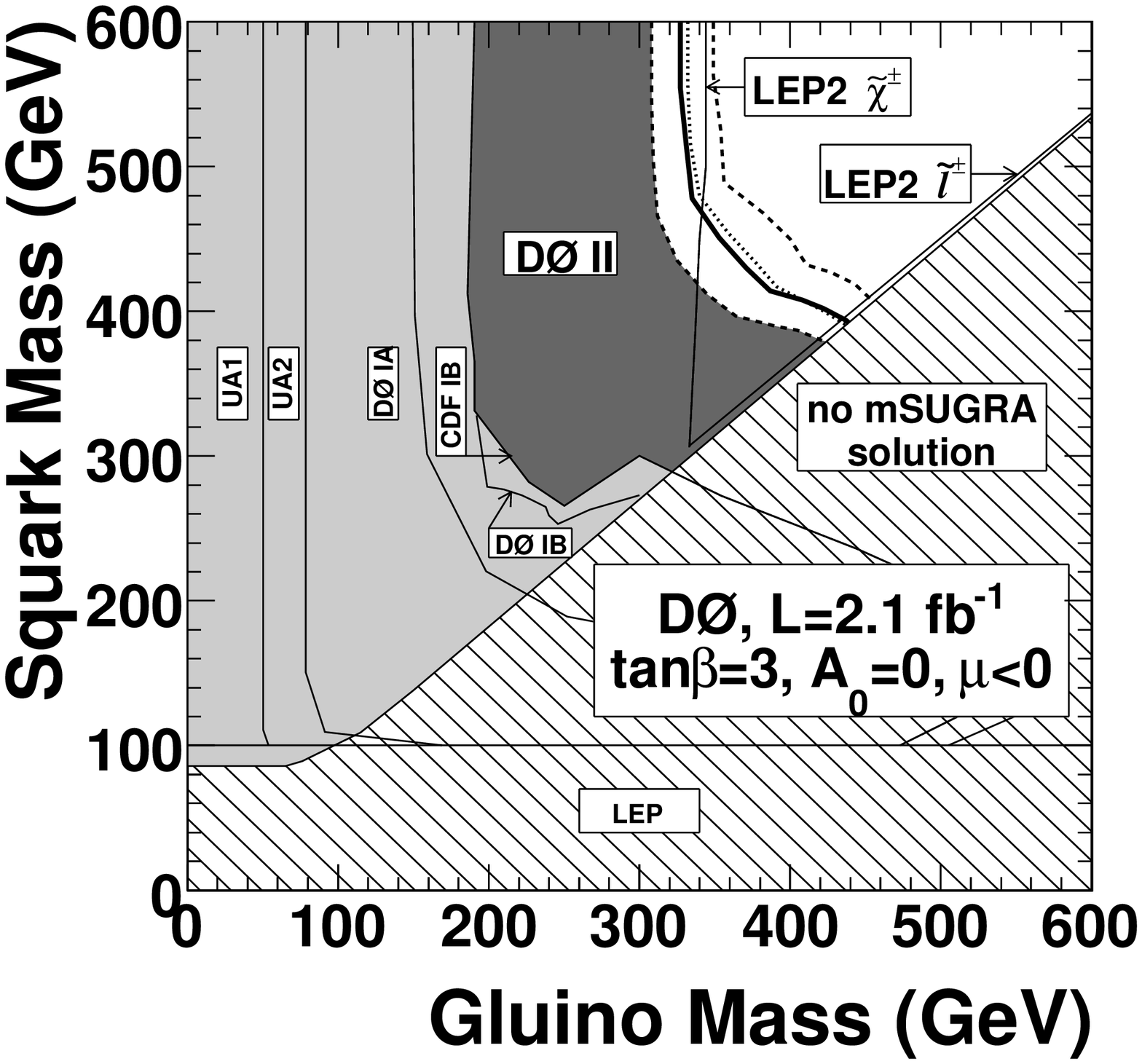}
\caption{\label{contour}
In the gluino and squark mass plane, excluded regions at the 95\% C.L.
by direct searches in the mSUGRA framework with
$\tan\beta = 3$, $A_0 = 0$, $\mu < 0$. 
The region excluded by this analysis and previous D0 Run~II 
results~\cite{Abazov:2006bj} in the most conservative hypothesis ($\sigma_{\mathrm{min}}$)
is shown in dark shading.
The thick (dotted) line is the limit of the observed (expected) excluded region for the
$\sigma_{\mathrm{nom}}$ hypothesis.
The band delimited by the two dashed lines shows the effect of the PDF choice and of the variation of 
$\mu_{\mathrm{r,f}}$ by a factor of two. 
Regions excluded by previous experiments are indicated in light 
shading~\cite{prevexp}. 
The two thin lines indicate the indirect limits inferred from the LEP2 chargino
and slepton searches~\cite{lepindirect}.
The region where no mSUGRA solution can be found is shown hatched.
}
\end{figure}

The numbers of events observed in the data are in agreement with the SM background 
expectation in the three analyses. Therefore, an excluded domain in the gluino-squark
mass plane was determined as follows. The three analyses were run over signal MC samples generated
in the gluino-squark mass plane to compute signal efficiencies. 
Then, to take advantage of the different features of the three analyses, 
seven independent combinations of the three selections were defined as 
shown in Table III, which were combined in the limit computations. 
The number of events passing at least one of the three analyses is 31 while the SM expectation is 
$32.6\pm1.7$(stat.)$^{+9.0}_{-5.8}$(syst.) events.
Figure~\ref{xseclim} shows the 95\% C.L. observed and expected 
upper limits on squark-gluino production cross sections for the three benchmark 
scenarios. Figure~\ref{contour} shows the excluded domain in the gluino-squark mass plane.

The absolute lower limits on the squark and gluino masses obtained in the most conservative 
hypothesis, $\sigma_{\mathrm{min}}$, are 379~GeV and 308~GeV, respectively. The corresponding 
expected limits are 377\,GeV and 312\,GeV. Table~\ref{finallimits2} summarizes these absolute 
limits for the three signal cross section hypotheses.
Limits were also derived for the particular case $m_{\sq}=m_{\sg}$. For 
$\sigma_{\mathrm{min}}$, squark and gluino masses below 390\,GeV are excluded, while 
the expected limit is 390\,GeV. The observed limit becomes 408\,GeV for 
$\sigma_{\mathrm{nom}}$, and 427\,GeV for $\sigma_{\mathrm{max}}$.

\begin{table}
\renewcommand{\arraystretch}{1.2}
\begin{center}
\caption{\label{finallimits2}
Absolute lower limits at the 95\% C.L. on the squark and gluino masses (in GeV) as a function 
of the choice of signal cross section hypothesis as defined in the text. 
Numbers in parentheses correspond to the expected limits. These limits are valid for the 
mSUGRA parameters $\tan\beta = 3$, $A_0 = 0$, $\mu < 0$.
}
\begin{ruledtabular}
\begin{tabular}{lcc}
Hypothesis	   		&  Gluino mass 	& Squark mass \\
\hline
$\sigma_{\mathrm{min}}$ 	& 308 (312)	& 379 (377)  \\
$\sigma_{\mathrm{nom}}$ 	& 327 (332)	& 392 (391)  \\
$\sigma_{\mathrm{max}}$   	& 349 (354)	& 406 (404)  \\
\end{tabular}
\end{ruledtabular}
\end{center}
\end{table}

\begin{figure}
\includegraphics[width=8.5cm]{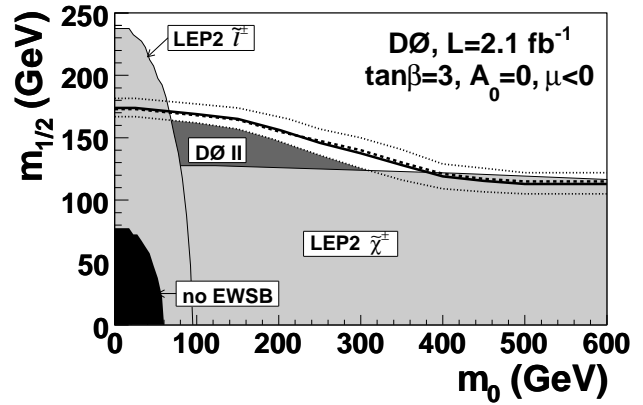}
\caption{\label{contour2}
In the $(m_{0},m_{1/2})$ plane, the region excluded by this analysis at the 95\% C.L.
in the mSUGRA framework for $\tan\beta = 3$, $A_0 = 0$, $\mu < 0$ is shown in dark shading.
The thick line is the limit of the excluded region for the $\sigma_{\mathrm{nom}}$ hypothesis.
The corresponding expected limit is the dashed line. The band delimited by the two dotted lines
shows the effect of the PDF choice and of the variation of $\mu_{\mathrm{r,f}}$ by a factor of two.
Regions excluded by the LEP2 chargino and slepton searches are indicated in light 
shading~\cite{lepindirect}.
The region where there is no electroweak symmetry breaking is shown in black.
}
\end{figure}

The results of this analysis also constrain the mSUGRA parameters at the
grand unification scale. Figure~\ref{contour2} shows the excluded regions in the
$(m_{0},m_{1/2})$ plane for $\tan\beta = 3$, $A_0 = 0$, $\mu < 0$.
Although a detailed scan of the mSUGRA parameter space is beyond the scope 
of this analysis, it was verified that similar results hold for a large 
class of parameter sets. In particular, the fact that there is no explicit veto
in this analysis on hadronically decaying tau leptons mitigates the expected
efficiency reduction at larger values of $\tan\beta$.
It can be seen in Fig.~\ref{contour2} that the limits from LEP2 chargino and slepton 
searches~\cite{lepindirect} are improved
for $m_0$ values between 70 and 300\,GeV and for $m_{1/2}$ values between 125 and 165\,GeV.
However, the LEP2 Higgs search limits remain more constraining 
in a purely mSUGRA scenario~\cite{lepindirect}.

In summary, a search for squarks and gluinos produced in $p\bar{p}$
collisions at 1.96\,TeV has been performed in a 2.1\,\invfb\ data sample. 
The results of three selections of events with jets 
and large missing transverse energy are in agreement
with the SM background predictions. 
In the framework of minimal supergravity with $\tan\beta = 3$, $A_0 = 0$,
and $\mu < 0$, 95\% C.L. lower limits of 392\,GeV and 327\,GeV were set on the 
squark and gluino masses, respectively, for the central choice of PDF and 
for a renormalization and factorization scale equal to the mass of the squark 
or gluino produced.
Taking into account the PDF uncertainties and allowing for
a factor of two in the choice of scale, these limits are reduced to 379\,GeV
and 308\,GeV. They exceed the corresponding previous limits~\cite{Abazov:2006bj} by 54\,GeV and 67\,GeV
and are the most constraining direct limits on the squark and gluino masses to date.

%
We thank the staffs at Fermilab and collaborating institutions, 
and acknowledge support from the 
DOE and NSF (USA);
CEA and CNRS/IN2P3 (France);
FASI, Rosatom and RFBR (Russia);
CAPES, CNPq, FAPERJ, FAPESP and FUNDUNESP (Brazil);
DAE and DST (India);
Colciencias (Colombia);
CONACyT (Mexico);
KRF and KOSEF (Korea);
CONICET and UBACyT (Argentina);
FOM (The Netherlands);
Science and Technology Facilities Council (United Kingdom);
MSMT and GACR (Czech Republic);
CRC Program, CFI, NSERC and WestGrid Project (Canada);
BMBF and DFG (Germany);
SFI (Ireland);
The Swedish Research Council (Sweden);
CAS and CNSF (China);
and the
Alexander von Humboldt Foundation.
%


\begin{thebibliography}{99}
%
\bibitem[a]{alton}
Visitor from Augustana College, Sioux Falls, SD, USA.
\bibitem[b]{burdin}
Visitor from The University of Liverpool, Liverpool, UK.
\bibitem[c]{podesta-lerma}
Visitor from ICN-UNAM, Mexico City, Mexico.
\bibitem[d]{quadt,meyer}
Visitor from II. Physikalisches Institut, Georg-August-University, G{\"o}ttingen, Germany.
\bibitem[e]{voutilainen}
Visitor from Helsinki Institute of Physics, Helsinki, Finland.

\bibitem[\dag]{IntFellows}
Fermilab International Fellow.
\bibitem[\ddag]{deceased}
Deceased.

%
\vskip 0.25cm

\bibitem{susy}
  H.E. Haber and G.L. Kane, Phys.\ Rep.\ {\bf 117}, 75 (1985).
\bibitem{rparity}
  P. Fayet, Phys.\ Lett.\ B {\bf 69}, 489 (1977).
\bibitem{cosmo}
  J. Ellis {\sl et al.}, Nucl.\ Phys.\ {\bf B238}, 453 (1984).

\bibitem{Abazov:2006bj}
  V.M.~Abazov {\sl et al.} (D0 Collaboration),
  Phys.\ Lett.\  B {\bf 638}, 119 (2006).

\bibitem{msugra}
  H.P. Nilles, Phys.\ Rep.\ {\bf 110}, 1 (1984).

\bibitem{Abazov:2005pn}	
  V.M.~Abazov {\sl et al.} (D0 Collaboration),
  Nucl.\ Instrum.\ Methods in Phys. Res. A {\bf 565}, 463 (2006).

\bibitem{jetalgo}
  G.C. Blazey {\sl et al.}, in {\sl Proceedings of the Workhop: ``QCD and Weak Boson Physics in Run II,''}
  edited by U.~Baur, R.K.~Ellis, and D. Zeppenfeld (Fermilab, Batavia, IL, 2000), p.\,47; see Sec. 3.5 
  for details.

\bibitem{Abolins:2007yz}
  M.~Abolins {\sl et al.},
  Nucl.\ Instrum.\ Methods in Phys. Res. A {\bf 584/1}, 75 (2007).

\bibitem{thomas}
  T.~Millet,
  PhD Thesis, unpublished,
  FERMILAB-THESIS-2007-22, LYCEN-T-2007-09.

\bibitem{geant} 
  R. Brun and F. Carminati, CERN Program Library Long Writeup W5013, 1993 (unpublished).

\bibitem{cteq6}
  J.~Pumplin {\sl et al.},
  JHEP {\bf 0207}, 012 (2002); 
  D.~Stump {\sl et al.},
  JHEP {\bf 0310}, 046 (2003).

\bibitem{Mangano:2002ea}
  M.L.~Mangano {\sl et al.},
  JHEP {\bf 0307}, 001 (2003);
  version 2.05 (2.11) was used for Run~IIa (Run~IIb) simulation.

\bibitem{Sjostrand:2006za}
  T.~Sj\"ostrand, S.~Mrenna and P.~Skands,
  JHEP {\bf 0605}, 026 (2006);
  version 6.323 (6.409) was used for Run~IIa (Run~IIb) simulation.

\bibitem{Boos:2004kh}
  E.~Boos {\sl et al.} (CompHEP Collaboration),
  Nucl.\ Instrum.\ Methods in Phys. Res. A {\bf 534}, 250 (2004).

\bibitem{Campbell:2001ik}
  J.M.~Campbell and R.K.~Ellis, 
  Phys.\ Rev.\ D {\bf 60}, 113006 (1999).

\bibitem{Djouadi:2002ze}
  A.~Djouadi, J.L.~Kneur and G.~Moultaka,
  Comput.\ Phys.\ Commun.\ {\bf 176}, 426 (2007).

\bibitem{Muhlleitner:2003vg}
  M.~Muhlleitner, A.~Djouadi and Y.~Mambrini,
  Comput.\ Phys.\ Commun.\ {\bf 168}, 46 (2005).

\bibitem{Beenakker:1996ch}
  W.~Beenakker, R.~Hopker, M.~Spira and P.M.~Zerwas,
  Nucl.\ Phys.\ {\bf B492}, 51 (1997).

\bibitem{vtxreco}
  V.M.~Abazov {\sl et al.} (D0 Collaboration),
  Phys.\ Rev.\ D {\bf 74}, 112004 (2006).

\bibitem{CLS}
  T. Junk, Nucl. Instrum. Methods in Phys. Res. A {\bf 434}, 435 (1999);
  A.~Read, in {\sl ``1st Workshop on Confidence Limits,''} CERN Report No. CERN-2000-005, 2000.

\bibitem{d0lumi}
  T.~Andeen {\sl et al.}, FERMILAB-TM-2365 (2007).

\bibitem{prevexp}
  C.~Albajar {\sl et al.}  (UA1 Collaboration),
  Phys.\ Lett.\ B {\bf 198}, 261 (1987);
  J.~Alitti {\sl et al.}  (UA2 Collaboration),
  Phys.\ Lett.\ B {\bf 235}, 363 (1990);
  S.~Abachi {\sl et al.}  (D0 Collaboration),
  Phys.\ Rev.\ Lett.\  {\bf 75}, 618 (1995);
  B.~Abbott {\sl et al.} (D0 Collaboration),
  Phys.\ Rev.\ Lett.\  {\bf 83}, 4937 (1999);
  T.~Affolder {\sl et al.}  (CDF Collaboration),
  Phys.\ Rev.\ Lett.\  {\bf 88}, 041801 (2002);
  A.~Heister {\sl et al.}  (ALEPH Collaboration),
  Phys.\ Lett.\ B {\bf 537}, 5 (2002);
  P.~Achard {\sl et al.}  (L3 Collaboration),
  Phys.\ Lett.\ B {\bf 580}, 37 (2004).

\bibitem{lepindirect}
  LEPSUSYWG, ALEPH, DELPHI, L3 and OPAL collaborations, note LEPSUSYWG/02-06.2
  (http://lepsusy.web.cern.ch/lepsusy/).

\end{thebibliography}
\end{document}